\documentclass[%
reprint,
showpacs,
amsmath,amssymb,
aps,prl,
]{revtex4-1}
\usepackage{booktabs}
\usepackage{tabularx}
\usepackage{braket}
\usepackage{float}
\usepackage{graphicx}
\usepackage{dcolumn}
\usepackage{bm}
\usepackage{subfigure}
\usepackage{color}
\usepackage{siunitx}

\usepackage[colorlinks,citecolor = blue, linkcolor=red,hyperindex,CJKbookmarks]{hyperref}

\usepackage{xcolor}

\newcommand{\figrefs}[1]{\mbox{Fig.\ref{#1}}}
\newcommand{\figrefm}[2]{\mbox{Fig.\ref{#1}#2}}

\newcommand\eg{e.g.\ }

\def\be{\begin{equation}}
\def\ee{\end{equation}}
\newcommand{\ba}{\begin{eqnarray}}
\newcommand{\ea}{\end{eqnarray}}

\definecolor{pink1}{RGB}{255,192,203}
\definecolor{pink2}{RGB}{255,130,180}
\definecolor{blue2}{RGB}{135,206,235}
\definecolor{blue1}{RGB}{176,224,230}

\newcommand{\bea}{\begin{eqnarray}}
\newcommand{\eea}{\end{eqnarray}}

\begin{document}

\title{Extended lifetime of respiratory droplets in a turbulent vapour puff and its implications on airborne disease transmission} 
\author{Kai Leong Chong$^{1}$}
\author{Chong Shen Ng$^{1}$}
\author{Naoki Hori$^{1}$}
\author{Rui Yang$^{1}$}
\author{Roberto Verzicco$^{1,2,3}$}
\author{Detlef Lohse$^{1,4}$}\email{d.lohse@utwente.nl}

\affiliation{$^1$Physics of Fluids Group and Max Planck Center for Complex Fluid Dynamics, MESA+ Institute and J. M. Burgers Centre for Fluid Dynamics, University of Twente, P.O. Box 217, 7500AE Enschede, The Netherlands\\
$^2$Dipartimento di Ingegneria Industriale, University of Rome `Tor Vergata', Roma 00133, Italy\\
$^3$Gran Sasso Science Institute - Viale F. Crispi, 7 67100 L'Aquila, Italy\\
$^4$Max Planck Institute for Dynamics and Self-Organization, 37077 G\"ottingen, Germany}

\date{\today}

\begin{abstract}
To quantify the fate of respiratory droplets under different ambient relative humidities, direct numerical simulations of a typical respiratory event are performed. We found that, because small droplets (with initial diameter of \SI{10}{\micro \meter}) are swept by turbulent eddies in the expelled humid puff, their lifetime gets extended by a factor of more than 30 times as compared to what is suggested by the classical picture by William F. Wells, for 50\% relative humidity. With increasing ambient relative humidity the extension of the lifetimes of the small droplets further increases and goes up to  around {150} times for {90}\% relative humidity, implying more than two meters advection range of the respiratory droplets within one second. Employing Lagrangian statistics, we demonstrate that the turbulent humid respiratory puff engulfs the small droplets, leading to many orders of magnitude increase in their lifetimes, implying that they can be transported much further during the respiratory events than the large ones. Our findings provide the starting points for larger parameter studies and may be instructive for developing strategies on optimizing ventilation and indoor humidity control. Such strategies are key in mitigating the COVID-19 pandemic in the present autumn and upcoming winter.
\end{abstract}

\maketitle
Tiny saliva and mucus droplets play a crucial role in the transmission of the disease SARS-CoV-2 \cite{asadi2020,bourouiba2020,mittal2020,setti2020,anderson2020,stadnytskyi2020,schijven2020,bahl2020,jayaweera2020,abkarian2020stretching,morawska2020,leung2020,wilson2020}. Hitherto laboratory studies have focused on the virological side, by investigating the viral load of the droplets \cite{wolfel2020virological}. Unfortunately, detailed knowledge on the realistic fates of respiratory droplets once they have been expelled is surprisingly sparse. Past studies have been limited in their ability to fully capture the turbulent flows that transport droplets, usually resorting to models \cite{wells1934,Xie2007}.
Such detailed knowledge, however, is vital to reduce the number of infections and the reproduction factor R of COVID-19. Whilst transmission of respiratory diseases depends on numerous factors, including infectivity and transmissability of the pathogen, one key question intimately related to the \textit{mode} of transmission, is that of fluid dynamics and flow physics: How do the turbulent vapour puff and ambient conditions influence the evaporation rate and thus the lifetime of the respiratory droplets?

The understanding of the fate of respiratory droplets is based on the classical picture by William F.\ Wells in 1930s \cite{wells1934,wells1936}, which -- at that time in connection with the transmission of tuberculosis -- was the following: The drops produced by sneezing, coughing and even speaking would have a wide size distribution and would fly out of the mouth and nose without much interaction between them. The small droplets would hardly be a problem because they would evaporate very quickly in the air and leave dry and therefore -- as was thought -- less dangerous airborne particles behind, while the large droplets would behave ballistically. To-date, on the basis of the classical picture by Wells and early experimental studies \cite{bahl2020}, authorities have implemented the social distance guidelines such as the \textit{six-foot rule}, to reduce the spread of COVID-19.

However, in recent months the empirical evidence that the six-foot rule is not sufficient to protect against infection with the coronavirus has kept on accumulating and various so-called super-spreading events have been reported, see e.g.\ 
\cite{miller2020,buonanno2020,guenther2020,prather2020,lu2020,lin2020,mat2020,NIID2020,bontempi2020,lelieveld2020}, all of them indoors. Indeed, over the last years L. Bourouiba and coworkers have shown \cite{bourouiba2014,scharfman2016,bourouiba2020,bourouiba2021} that the  range and the lifetime of the cloud of tiny saliva and mucus droplets (referred to as respiratory droplets in this paper) is much larger than what the 6-foot rule assumes, namely up to 8 meters and up to 10 minutes, instead of one to two meters and less than one second. The reason justified by these studies is that the respiratory droplets are expelled together with warm and humid air, which considerably delays their evaporation. 
In a dense spray, Villermaux and coworkers \cite{derivas2016,villermaux2017} pointed out that the lifetime of droplets is determined by the lifetime of the humid vapour plume in which the droplets are embedded, rather than by the so-called ``$d^2$-law'' which only applies to an {\it isolated} evaporating spherical droplet, whose square of its diameter $d$ linearly decreases with time \cite{langmuir1918} and on which the estimates by Wells \cite{wells1934,wells1936} were based. 

\begin{figure}
	\includegraphics[width=1.0\columnwidth]{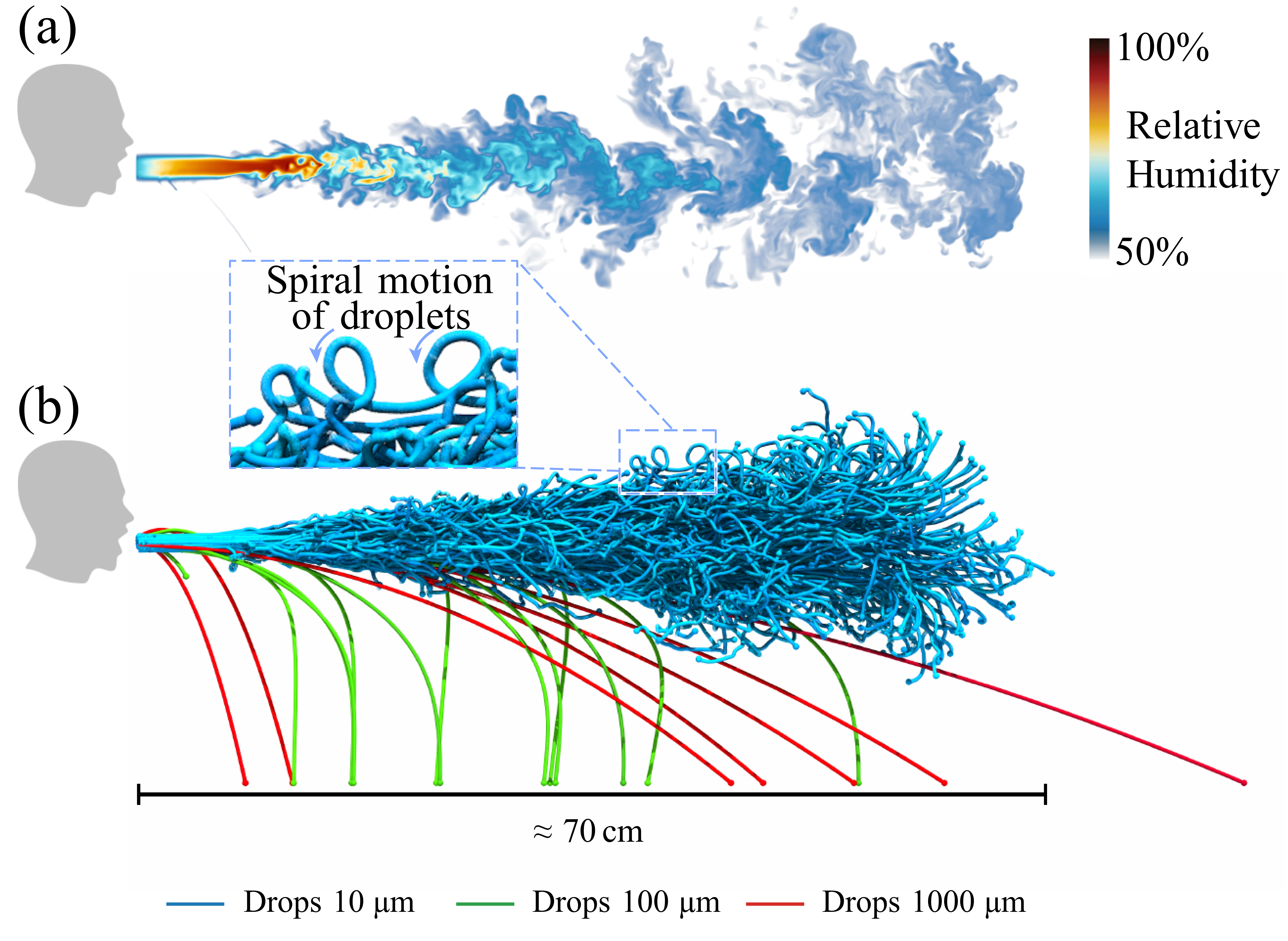}
	\caption
	{Snapshot of the droplet-laden cough simulation, (a) contour of the relative humidity field, and (b) droplet trajectories at time $t=\SI{400}{\milli \second}$. Trajectories of three different diameters are shown: $\SI{10}{\micro \meter}$ (blue), $\SI{100}{\micro \meter}$ (green), and $\SI{1000}{\micro \meter}$ (red), respectively. Larger droplets are observed to fall out from the puff whereas smaller droplets remain protected and are carried along by the puff. For discussion purposes, only a subset of the total number of droplets are shown.}
	\label{fig:coughimages}
\end{figure}

Interestingly, in spite of the major research effort of the last months in particular, the role of environmental factors such as humidity and temperature remains controversial and inconclusive \cite{eslami2020,kumar2020,Mecenas2020}, motivating several studies to focus on mathematical models. The main difficulty in getting conclusive and reproducible results originates from the lack of controlled conditions under which the spreading events occur. Also, even with controlled and reproducible conditions, following and quantifying the properties of 1000s of microdroplets in space and time remains extremely challenging from an experimental point-of-view. High fidelity numerical datasets are thus crucial in order to pinpoint the exact flow physics that determine the droplet evaporation.

In this Letter, we perform direct numerical simulations (DNS) of a turbulent respiratory event and follow the evolution of microdroplets in turbulent flow using the point particle approach. In contrast to the classical picture by Wells, we show that the lifetime of droplets can be significantly extended in the presence of the turbulent vapour puff, which we quantify, for the first time, by tracking the Lagrangian statistics of droplets in our DNS. Existing numerical studies on this topic, moreover, typically do not resolve small scales of turbulent mixing processes--a crucial mechanism for the droplet evaporation. Examples of these studies include Euler-Lagrangian approaches with a steady-state jet profile for a single droplet \cite{Xie2007,liu2017}, for multiple droplets \cite{busco2020,feng2020,dbouk2020coughing} and large eddy simulations (LES) \cite{abkarian2020}. Mathematical models are appealing because of their simplicity \cite{marr2019mechanistic,chaudhuri2020,balachandar2020host,mittal2020mathematical}, however, fitting parameters must be calibrated by validating with high fidelity simulations and experiments. Newer DNS with turbulence are emerging \cite{diwan2020,rosti2020virusfate}, but full quantification of the Lagrangian statistics is still limited. Therefore, the quantitative results obtained here are an important step towards developing a theoretical model on predicting spatial and temporal droplet or aerosol concentrations around a respiratory event.

\begin{figure}
\includegraphics[width=1.0\columnwidth]{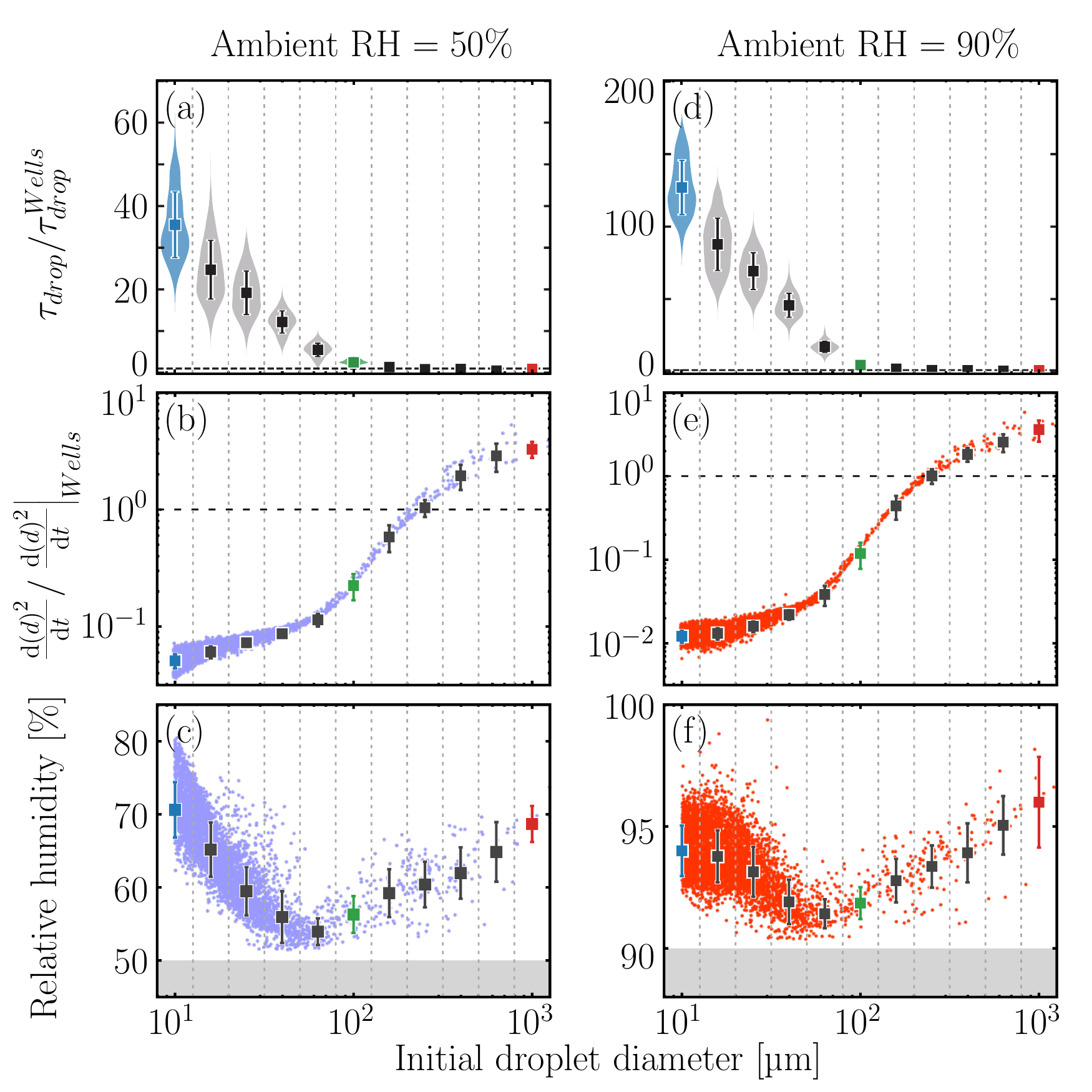}
\caption
{Lagrangian statistics of droplets plotted against the droplet initial diameters for (a-c) ambient RH = 50\% and (d-f) ambient RH = 90\%. (a,d) Lifetime distribution of droplets compared to lifetime estimation from Wells \cite{wells1934} (dashed horizontal line).  (b,e) Averaged change of the surface area of a droplet throughout its lifetime compared to that from Wells' estimation (dashed horizontal line). (c,f) Averaged relative humidity. The different coloured symbols denote different initial droplet diameters: $\SI{10}{\micro \meter}$ (blue), $\SI{100}{\micro \meter}$ (green), and $\SI{1000}{\micro \meter}$ (red), as shown in \figrefm{fig:coughimages}{b}. Plots show mean values with one standard deviation ranges.}
\label{fig:coughdroplets}
\end{figure}

We simulated a turbulent humid puff sustained over a duration of \SI{0.6}{\second} into ambient air and laden with {5000} 
water droplets, mimicking a strong jet-like cough \cite{tang2013airflow,bourouiba2014}. In addition to the droplets, the turbulent puff expels hot, vapor saturated air with an initial temperature $\SI{34}{\degreeCelsius}$ and relative humidity $100$\% \cite{bourouiba2014}. Both temperature and vapor fields are buoyant. In reality, no two coughs are alike, and deviations from our assumed parameters can exist. Here, to make our simulations tractable, we have chosen parameters that are representative of a cough from experiments \cite{gupta2009flow,tang2009schlieren}. As reference ambient conditions, we chose the ambient temperature of  $\SI{20}{\degreeCelsius}$ and varied relative humidity between 50\% and 90\%, covering typical indoor ambient conditions. The background airflow conditions can also be an important factor, \eg\,with or without ventilation \cite{somsen2020}. Here we chose a quiescent background field, but different types of ventilation could be embodied straightforwardly. We use our highly efficient and parallelized finite difference Navier-Stokes solver (``AFiD'') based on a second-order finite difference scheme \cite{vanderpoel2015}, coupled to the advection equations for temperature and vapor concentration, both in Boussinesq approximation \cite{yang2016pnas}. Details of the numerical scheme and setup are given in the Supplementary Material.
 
We first describe our results for fixed ambient relative humidity of RH = 50\%, visualized in \figrefm{fig:coughimages}. Within about \SI{400}{\milli \second} from the start of the cough, droplets larger than about \SI{100}{\micro \meter} are observed to immediately fall out from the `puff', basically behaving ballistically due to their own weight. The associated distances of this fallouts typically range between \SI{0.1}{\meter} and \SI{0.7}{\meter} from the source (\figrefm{fig:coughimages}{b}). Indeed, this type of fallout has already been predicted in the 1930s by Wells \cite{wells1934,wells1936}, and also demonstrated in the cough and sneeze experiments by \cite{bourouiba2014}. These typical distances appear to be the basis of the spatial separation guidelines issued by the World Health Organization (WHO), Centers for Disease Control and Prevention, and European Centre for Disease Prevention and Control on respiratory protection for COVID-19 \cite{bahl2020}.

However, droplets of order of \SI{10}{\micro \meter} behave in a completely different way. The pathlines traced by smaller droplets remain largely horizontal and form spirals, indicating close correlation with the turbulent puff. The physical explanation here is that smaller droplets settle much slower than the characteristic velocity of the surrounding fluid, and are therefore advected further by the local turbulent flow. This mechanism is intimately related to the `airborne' transmission route for infectious diseases \cite{jones2015}. 

That the smaller droplets tend to remain inside the humid puff has dramatic consequences on their \textit{lifetimes}, which far exceed those of isolated droplets (\figrefm{fig:coughdroplets}{a,d}). For RH = 50\%, the droplets of \SI{10}{\micro \meter} can live up to 60 times longer than the expected value by Wells, whereas for RH = 90\%, it can even become $100$ to $200$ times longer. These extended lifetimes are also confirmed by the much slower shrinkage rates of the droplet surface area as compared to the corresponding rate determined from the $d^2$-law,  which is valid for isolated droplets and which is the basis for Wells' theory, see \figrefm{fig:coughdroplets}{b,e}.

We now further describe the flow physics contributing to this highly extended lifetime of the small droplets. The first physical factor depends on the motion of droplets relative to their surrounding fluid. As shown in \figrefm{fig:coughimages}{b}, smaller droplets have the tendency to be captured by the turbulent puff and move together with the fluid. This is the well-known phenomenon for small Stokes number particles, defined as a ratio of droplet response timescale to the flow timescale. This gives rise to smaller relative velocities and hence less evaporation due to the reduction of convective effects. In contrast, larger droplets tend to fly and settle faster than the surrounding fluid, thus evaporating faster than predicted by the $d^2$-law,  because the convective effects, carrying the evaporated vapor away from the droplet, are dominant. This rapid evaporation is shown in the faster shrinkage rate of the droplet surface areas in \figrefm{fig:coughdroplets}{b,e}.

The second and more crucial factor is the influence of the humid air around the small droplets, originating from the humid puff and the ambient surroundings. In order to quantify this effect on the lifetime, we show the averaged relative humidity of the droplets throughout the simulation as function of the initial droplet diameter $d$ in \figrefm{fig:coughdroplets}{c,f}. From this figure, we observe a clear non-monotonic behavior, reflecting two  different regimes. In the first regime for small droplets $d = 10 - 100 \si{\micro \meter}$, the relative humidity takes higher values than the ambient, reflecting that the droplets are surrounded by nearly saturated humid vapor. As the initial droplet size increases, the relative humidity decreases because the settling speed increases and the droplets tend to stray from the puff (\figrefm{fig:coughimages}{b}). In the second regime, for large droplets $d > \SI{100}{\micro \meter}$, however, we observe that the relative humidity increases with the increase in size. The reason is that larger droplets evaporate larger quantities of vapor per given time, which leads to higher relative humidity in their surroundings. This effect is very local, due to the strong shear around the falling droplets. 

\begin{figure}
\includegraphics[width=1.0\columnwidth]{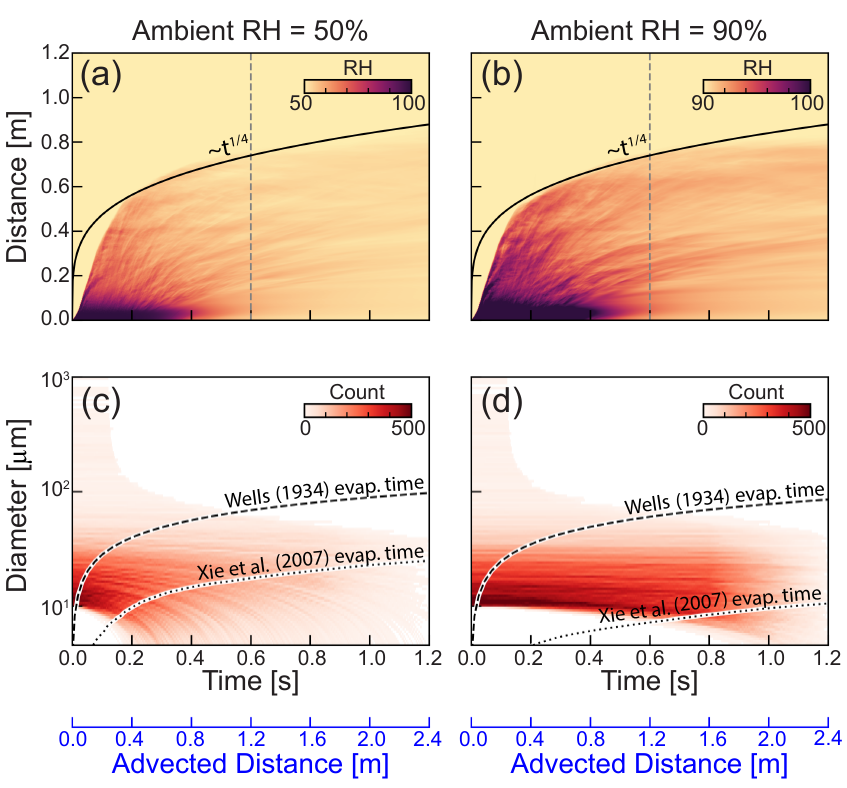}
\caption
{(a,b) Relative humidity variation in space and time. The dark violet colour indicates the region with high relative humidity which can protect the droplets from shrinking. The vertical dashed line indicates the moment at which the coughing stops. (c,d) Count histogram of droplets in the entire domain at a given size and time. Time is shifted to the expulsion time for each droplet. The dashed line delineates the expected droplet lifetimes that completely evaporate, which is computed according to the assumptions by Wells \cite{wells1934}, i.e., based on the $d^2$-law, at matched RH values. The dotted lines are the expected droplet lifetime reproduced from Xie \textit{et al.} \cite{Xie2007} for matched conditions. The blue axes below panels (c) and (d) show estimated advected distance of small droplets based on a background advecting velocity of \SI{2}{\meter\per\second}.}
\label{fig:RHspacetime}
\end{figure}

\begin{figure*}
\includegraphics[width=1.9\columnwidth]{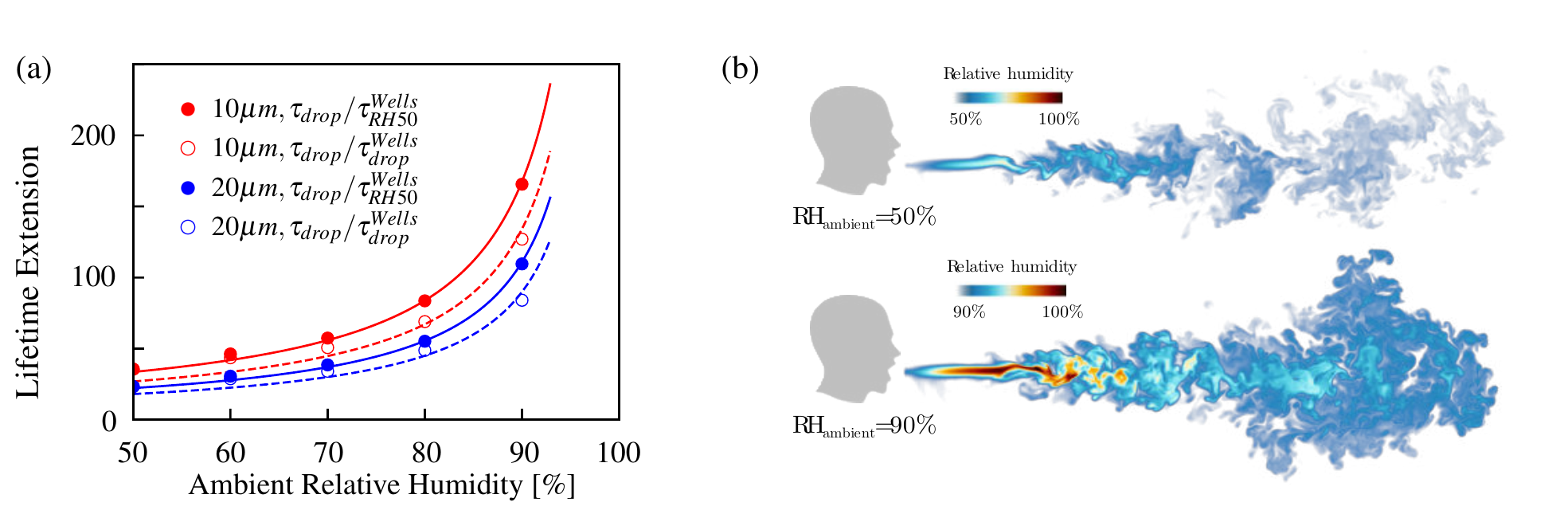}
\caption
{(a) Extended lifetime as a function of relative humidity up to RH = 90\%. The curves in the figure are fitted according to the function $y=a_1/(1-x)+a_2$, where $a_1$ and $a_2$ are the fitting parameters. (b) Contour  of  the  relative  humidity  fields for ambient RH = 50\% and 90\% at time \SI{600}{\milli \second}.}
\label{fig:ExtendedDropletLifetime}
\end{figure*}

Since the humid puff leads to extended droplet lifetimes, it is instructive to examine the propagation of the humidity field. Indeed, also in the case of dense sprays \cite{derivas2016,villermaux2017}, the droplets' fate is determined by the vapor concentration field. In \figrefm{fig:RHspacetime}{a,b}, we show the relative humidity in the puff as function of time and distance from the respiratory release event, in order to quantify the propagation of the puff after exhalation. Very soon after the respiratory event starts, moist air coming out of the mouth creates very high humidity region in the vicinity $\approx \SI{0.3}{\meter}$ of the mouth in \SI{0.2}{\second}. The overall humidity rapidly decreases after the respiratory event stops, and the puff continues to propagate because of the conservation of momentum with the puff edge growing proportionally as $t^{1/4}$
\cite{bourouiba2014}, see the solid line in \figrefs{fig:RHspacetime}, which is an upper bound to our numerical results. Note that this transition to $t^{1/4}$ is also observed in speech experiments \cite{abkarian2020}. There, the transition time and distance exhibit shifts due to small differences in how words are pronounced. Additional simulations with different mouth shapes and speeds in Supplemental Material \cite{prlsupp} also confirm this and show the generality of our results.

The extended lifetime of the small droplets can also be expressed in the so-called evaporation-falling curve, as introduced in the classical work by  William F. Wells \cite{wells1934}. He derived the dependence of the  lifetime of the droplet on its  size, based on the $d^2$-law for evaporation of an isolated droplet and on the droplet settling time, see the dashed curves in \figrefm{fig:RHspacetime}{c,d}. According to this classical theory by Wells, droplets below this line evaporate completely and thus should not exist. In those figures we now include the histograms of the counts of droplets at given size and time, based on our DNS at RH of 50\% and 90\%, respectively. Note that the time (on the $x$-axis) is shifted to the expulsion time for each droplet, such that the history of each droplet begins at time = \SI{0}{\second}. \figrefm{fig:RHspacetime}{c,d} clearly show that a considerable number of small droplets do exist below the classical Wells curve, demonstrating that the classical Wells estimate is inappropriate and that small droplets can live much longer. In fact, Xie \textit{et al.} \cite{Xie2007} have already addressed the limitations in the study of Wells, namely by coupling temperature and humidity fields to a steady-state jet and assuming single droplet. For comparison, we plot the curves from Xie \textit{et al.} in \figrefm{fig:RHspacetime}{c,d}. The small droplets from our simulations can even exceed the Xie \textit{et al.} estimates. If a background flow of \SI{2}{\meter/\second} exists (nominal wind speed at acceptable comfort levels \cite{willemsen2007design}), these long-lived droplets can easily travel much farther than \SI{2}{\meter} in a second as carried by the turbulent puff, as shown in the advected distances estimated in \figrefm{fig:RHspacetime}{c,d}. 

To study the dependence of different RH, we repeat our DNS for ambient RH values in the window 50\% $\le$  RH  $\le$  90\%. From \figrefm{fig:ExtendedDropletLifetime}{a}
one can observe that the lifetime of the small droplets increases dramatically and even diverges to infinity at RH = 100\%. The smaller the droplets, the more pronounced the effect. For the smallest respiratory droplets of this study with initial diameter $d=$ \SI{10}{\micro \meter}, for RH = 90\%  the lifetime extension is with a factor of about 130 as compared to the lifetime of a droplet behaving according to the Wells model, and even with a factor of 166 as compared to the lifetime of Wells model droplet at RH = 50\%. Similarly, for slightly larger droplets with $d=$ \SI{20}{\micro \meter}, the lifetime extension remains significant with a factor of 80 and 110, respectively. The first reason for the dramatic increases in droplet lifetime seen in \figrefm{fig:ExtendedDropletLifetime}{a} is a significantly reduced evaporation rate for larger ambient RH as the ambient gas is much closer to the saturated condition. The second reason is that for larger ambient RH the vapour puff can be sustained for longer times and distances, as shown in \figrefm{fig:ExtendedDropletLifetime}{b}. In such case, there is stronger protection from the vapour puff for larger ambient RH.

In conclusion, our DNS results are consistent with the multiphase cloud emission model \cite{bourouiba2020}, but are inconsistent with Wells' classical model \cite{wells1934,wells1936}. The reason is that Wells' model assumed that the droplets are isolated, i.e., have no interaction with the near velocity, temperature, and humidity fields around the droplet, which is far from the case in reality. Indeed, our study has conveyed that in particular the humid vapor exhaled together with the droplets must not be neglected, as the vapor concentration around the droplet remains high during the whole  respiratory event and thereafter (see \figrefm{fig:coughdroplets}{a,d}), strongly contributing to the lifetime extension of the small droplets by orders of magnitude. In this sense the lifetime of the respiratory droplets is mainly controlled by the mixing \cite{villermaux2019} of the humidity field exhaled together with them, similarly as occurring for the lifetime of evaporating dense sprays, which is also controlled by the mixing of the vapor field \cite{derivas2016,villermaux2017}. The relevant length scale for droplet evaporation is therefore not the diameter of the droplet itself (sub-millimeter), but the outer length scale of the surrounding turbulent velocity and humidity field, i.e., meters. 

The lifetime of respiratory droplets has, to-date, grossly been underestimated (see \figrefs{fig:RHspacetime}). The extension of the droplet lifetime is so extreme that the  smallest  droplets ($d=$\SI{10}{\micro \meter}) of our study barely evaporate (lifetime extension by factor 35 at ambient relative humidity RH = 50\% and even by factor 166 at RH = 90\%) and are transported in an aerosolised manner. This finding contradicts the 'respiratory droplet' classification by WHO for $d>$5-\SI{10}{\micro \meter} droplets \cite{world2020transmission}, which implies that droplets of these sizes fall ballistically. From our results, there is strong evidence that even smaller droplets with initial diameter $d\approx$ \SI{10}{\micro \meter} will survive even far longer, because of the protection from the turbulent humid puff. From \figrefs{fig:coughdroplets}, one can estimate the correction factor for the droplet lifetime for given droplet size, as compared to the Wells estimate. However, in the present study only the coughing case has been studied. In future work, more studies for other respiratory events with different Reynolds numbers and ambient conditions should be conducted to fully parametrize the correction factor. For example, for sneezing, while the expulsion velocity of droplets is larger, shorter lifetime of droplets is expected due to faster dissipation of the vapour puff by stronger entrainment.

Our results also show that the lifetime extension of the respiratory droplets is more significant when the ambient RH is higher, see \figrefm{fig:ExtendedDropletLifetime}{a}. The reason lies in the longer lifetime of the local humidity field around the droplets for larger ambient RH, consistent with the picture that the mixing of the local humidity field determines the droplet lifetime, see \figrefm{fig:ExtendedDropletLifetime}{b}. Indeed, L. Bourouiba and coworkers \cite{bourouiba2014,scharfman2016,bourouiba2020,bourouiba2021} have highlighted the qualitative effect of the vapour puff on the longer airborne transmission of droplets in their experiment. Here, with the help of Lagrangian statistics, we have quantified this lifetime extension and further elucidated under what conditions and why it occurs. This finding has important ramifications on the transport of droplets in case of high RH, which has implications on the spread of COVID-19. This finding is also informative for optimizing mitigation strategies such as controlling indoor humidity and ventilation \cite{bhagat2020effects}. Finally, we emphasize that we have only focused on the flow physics of respiratory droplets, and our conclusion does not directly infer on the infectivity of virus-laden droplets or droplet nuclei.

{\it Acknowledgements:}  
This work was funded by the Netherlands Organisation for Health Research and Development (ZonMW), project number 10430012010022: Measuring, understanding \& reducing respiratory droplet spreading", the ERC Advanced Grant DDD, No. 740479, and Foundation for Fundamental Research on Matter with project number 16DDS001, which is financially supported by the Netherlands Organisation for Scientific Research (NWO). The funders have no role in study design, data collection, and analysis or decision to publish. The simulations were performed on the national e-infrastructure of SURFsara, a subsidiary of SURF cooperation, the collaborative ICT organization for Dutch education and research, the DECI resource Kay based in Ireland at the Irish Center for High-End Computing (ICHEC) with support from the PRACE aisbl, and Irene at Très Grand Centre de calcul du CEA (TGCC) under PRACE project No. 2019215098.

K.L.C. and C.S.N. contributed equally to this work.


\end{document}